\batchmode
\makeatletter
\def\input@path{{\string"C:/Documents and Settings/BTI/My Documents/References/Photons/Photon Models/Bateman/BatemanOnLight/\string"/}}
\makeatother
\documentclass[oneside,english,reqno]{amsart}
\usepackage[T1]{fontenc}
\usepackage[latin1]{inputenc}
\usepackage{geometry}
\usepackage{setspace}
\usepackage{amssymb}

\makeatletter
 \theoremstyle{plain}

\usepackage{babel}
\makeatother
\begin{document}

\title{Bateman On Light: Synchrotron Rays}

\author{H. C. Potter}

\keywords{Light, Duality, Localization, Singularities, Potentials, Fields}

\begin{abstract}
For over a century physicists have sought a mathematical theory that
unites the dual, wave and corpuscle, behaviors that light exhibits.
In the early twentieth century H. Bateman was studying a solution
that uses wave functions with singularities to mark light quanta location
in space and time as potential singularities do for material particles.
He published an incomplete, mathematically obtuse paper applying his
scheme to presciently predict synchrotron rays well before they were
experimentally observed. Here, I clarify the presentation in Bateman's
paper and supply a missing demonstration that it predicts orbit plane
localization and previously unexplained orbit plane normal polarization. 
\end{abstract}
\maketitle
\newpage{}

In the twentieth century first quarter, H. Bateman (1882-1946) was
studying electromagnetic theory based on a generalized potential formalism.
This formalism is elaborated in his book \cite{Wave-motion}. Two
applications are significant. On cursory reading, the first \cite{Rays}
appears to predict synchrotron rays more than 30 years prior to their
observation \cite{Elder} and classical power spectrum prediction
\cite{Schwinger}. On the same basis, the second \cite{Pairs} appears
to predict pair production 9 years before its 1933 observation \cite{Blackett}
with no other extant electromagnetic explanation. I use the word {}``appears''
intentionally, because Bateman never fully developed these applications.
In this paper I apply the first paper to explain synchrotron light
polarization normal to the orbit plane and to predict ray localization
within the orbit plane.

Bateman takes as his generalized potential the quantity\begin{equation}
V=\frac{f(\tau)}{\sigma}\label{eq:GenPotential}\end{equation}
where\begin{equation}
\sigma=(x-\xi)l+(y-\eta)m+(z-\zeta)n-(t-\tau)c^{2}p\tag{B4}{}\label{eq:sigma}\end{equation}
and all the $l,m,n,p$ and $\xi,\eta,\zeta$ may be $\tau$ dependent
functions. The function $V$ will satisfy the wave equation\begin{equation}
\square_{c}V=\frac{\partial^{2}V}{\partial x^{2}}+\frac{\partial^{2}V}{\partial y^{2}}+\frac{\partial^{2}V}{\partial z^{2}}-\frac{1}{c^{2}}\frac{\partial^{2}V}{\partial t{}^{2}}=0\label{eq:daLemV}\end{equation}
when $\sigma\neq0$ and\begin{equation}
l^{2}+m^{2}+n^{2}=c^{2}p^{2}.\tag{B3b}{}\label{eq:WaveCondition}\end{equation}

The function $\sigma$ defined by Eq.(\ref{eq:sigma}) vanishes when
$(x,y,z,t)$ lies on a curve $\Gamma$ defined by\begin{equation}
\begin{array}{ccccc}
x=\xi(\tau), & y=\eta(\tau), & z=\zeta(\tau) & \textnormal{}{and} & t=\tau.\end{array}\tag{B1}{}\label{eq:gamma}\end{equation}
 But $\sigma$ also vanishes when $(x,y,z,t)$ lies on lines through
$(\xi(\tau),\eta(\tau),\zeta(\tau),\tau)$ for $t>\tau$. To prove
this consider the sphere\begin{equation}
[x-\xi(\tau)]^{2}+[y-\eta(\tau)]^{2}+[z-\zeta(\tau)]^{2}=R^{2}.\tag{B2'}{}\label{eq:R^2sphere}\end{equation}
The Eq.(\ref{eq:R^2sphere}) sphere represents a retarded wave front
when\begin{equation}
\tau=t-\frac{R}{c}.\label{eq:retardation}\end{equation}
 Eq.(\ref{eq:sigma}) will vanish at\begin{equation}
\begin{array}{ccccc}
x=R\frac{l}{cp}+\xi(\tau), & y=R\frac{m}{cp}+\eta(\tau), & z=R\frac{n}{cp}+\zeta(\tau) & \textnormal{}{and} & t=\frac{R}{c}+\tau.\end{array}\label{eq:TangentPoint}\end{equation}
So when Eqs.(\ref{eq:WaveCondition}) and (\ref{eq:retardation})
are satisfied, Eq.(\ref{eq:TangentPoint}) parameter $R$ defines
a line on which $\sigma$ vanishes that passes through the Eq.(\ref{eq:R^2sphere})
sphere center at $(\xi(\tau),\eta(\tau),\zeta(\tau),\tau)$. For fixed
$\tau$, Eq.(\ref{eq:retardation}) says the Eq.(\ref{eq:TangentPoint})
point moves with velocity $\mathbf{c}=(\frac{l}{p},\frac{m}{p},\frac{n}{p})$.
At $\tau$ on the curve $\Gamma$, the velocity is $\mathbf{v}=(\xi',\eta',\zeta')$
when the primes indicate $\tau$ differentiation. So, says Bateman,
\begin{equation}
l(\tau)\xi'(\tau)+m(\tau)\eta'(\tau)+n(\tau)\zeta'(\tau)=(-v^{2})p(\tau)\tag{B3a'}{}\label{eq:BatemanCondition}\end{equation}
determines the angle at the Eq.(\ref{eq:R^2sphere}) sphere center
between the singularity velocity on the Eq.(\ref{eq:TangentPoint})
line and the singularity velocity on the Eq.(\ref{eq:gamma}) curve
$\Gamma$. For plane, circular orbits like that described by Eq.(\ref{eq:ExampleGamma}),
below, Eq.(\ref{eq:BatemanCondition}) takes the Eq.(\ref{eq:SpawnCondition})
form; but there is no extant proof that Eq.(\ref{eq:BatemanCondition})
is valid for any other case. When $\mathbf{v}$ does not change direction,
this development will exhibit cylindrical symmetry about the line
defined by $\Gamma$. This is manifest in the plane orbit ray case
examples below where radial acceleration is ignored. When $\mathbf{v}$
does  change direction, this symmetry should be broken. The latter
contradicts common light beam depictions which presume the cylindrical
symmetry for linear motion to persist. The presumption is refuted,
however, by \cite[Fig. 5 frame 10 and Ref.2]{Elder}. This shows beam
size increase in the orbit plane with the energy sustaining r-f turned
off. Confounding by electron beam cross section convolution may explain
this error. 

As a mathematician, Bateman attempted to perform this development
using Eqs.(\ref{eq:gamma}), (\ref{eq:R^2sphere}), (\ref{eq:BatemanCondition}),
(\ref{eq:WaveCondition}) and (\ref{eq:sigma}) as initial premises.
In all Bateman equations with primed equation numbers, including Eq.(\ref{eq:PotentialDefs})
below, the parenthetical factors designate my revisions. I find Bateman's
development unintelligible, even though he does revisit this problem
in his book \cite[Sec. 44]{Wave-motion} without reference to the
earlier work. The development above, however, does produce Bateman's
result that for a primary Eq.(\ref{eq:GenPotential}) singularity
M moving with velocity $\mathbf{v}$ on a curve $\Gamma$ there are
at least two secondary Eq.(\ref{eq:GenPotential}) singularities radiating
with velocity $c$ along {}``rectilinear rays through each point
of this curve'' with the angle between the rays and the primary singularity
motion having a cosine equal to $-v/c$. This $v$ angle dependence
agrees with \cite[Eq. 1.48]{Schwinger} when expressed as his $\sin^{2}$
dependence. 

To connect the secondary singularities with light, Bateman defines
the vector and scalar potentials as\begin{equation}
\begin{array}{ccccc}
A_{x}=\frac{l}{\sigma}, & A_{y}=\frac{m}{\sigma}, & A_{z}=\frac{n}{\sigma} & \textnormal{}{and} & \Phi=\frac{(c)p}{\sigma}.\end{array}\tag{B5'}{}\label{eq:PotentialDefs}\end{equation}
When primary singularity velocity $v$ is always less than $c$, the
Eq.(\ref{eq:R^2sphere}) spheres for proximate instants are nested.
So the Eq.(\ref{eq:PotentialDefs}) potentials at $(x,y,z,t)$ will
be uniquely associated with a curve $\Gamma$ point for $v<c$. With
$\sigma$ defined by Eq.(\ref{eq:sigma}), Eq.(\ref{eq:WaveCondition})
causes the Eq.(\ref{eq:PotentialDefs}) potentials to satisfy the
Lorenz condition\begin{equation}
\boldsymbol{\nabla}\bullet\mathbf{A}+\frac{1}{c}\frac{\partial\Phi}{\partial t}=0,\label{eq:LorenzCondition}\end{equation}
where $\boldsymbol{\nabla}=(\frac{\partial}{\partial x},\frac{\partial}{\partial y},\frac{\partial}{\partial z})$.
When the electric and magnetic fields are calculated from the Eq.(\ref{eq:PotentialDefs})
potentials using the relations \cite[Eqs.(268)]{Wave-motion}\begin{equation}
\begin{array}{ccc}
\mathbf{H=\boldsymbol{\nabla}\times\mathbf{A}} & \textnormal{}{and} & \mathbf{E}=-\boldsymbol{\nabla}\Phi-\frac{1}{c}\frac{\partial\mathbf{A}}{\partial t}\end{array}\label{eq:FieldDefs}\end{equation}
with $v$ and the $l,m,n$ and $p$ constant on $\Gamma$, the fields
obtained are null.  When $v$ varies with $\tau$ however, Eq.(\ref{eq:BatemanCondition})
forces variation with $\tau$ for some or all the $l,m,n,p$ and $\xi',\eta',\zeta'$,
and the calculated fields need not be null. These fields satisfy the
Eq.(\ref{eq:daLemV}) wave equation, because the Eq.(\ref{eq:PotentialDefs})
potentials do; and they satisfy the vacuum Maxwell equations \cite[Eq.(1)]{Wave-motion}\begin{equation}
\left.\begin{array}{cc}
\boldsymbol{\nabla}\times\mathbf{H}=\frac{1}{c}\frac{\partial\mathbf{E}}{\partial t}, & \boldsymbol{\nabla}\times\mathbf{E}=-\frac{1}{c}\frac{\partial\mathbf{H}}{\partial t}\\
\boldsymbol{\nabla}\bullet\mathbf{H}=0, & \boldsymbol{\nabla}\bullet\mathbf{E}=0.\end{array}\right\} \label{eq:MaxwellEqs}\end{equation}

When some or all the $l,m,n,p$ and $\xi',\eta',\zeta'$ vary with
$\tau$ the above relations are satisfied at each $\tau$ value defining
a position on the Eq.(\ref{eq:gamma}) curve $\Gamma$. But the actual
fields spawned at $\tau$ onto the expanding Eq.(\ref{eq:R^2sphere})
sphere along the line given by Eq.(\ref{eq:TangentPoint}) can be
taken to be given by the Eqs.(\ref{eq:PotentialDefs}) potential changes
at $\tau$. The Eqs.(\ref{eq:FieldDefs})  magnetic and electric field
components spawned at $\tau$ are then given by\begin{equation}
\left.\begin{array}{ccccc}
H_{x} & = & \frac{\partial A_{z}}{\partial\eta}-\frac{\partial A_{y}}{\partial\zeta} & = & \frac{n'}{\sigma}\frac{\partial\tau}{\partial\eta}-\frac{m'}{\sigma}\frac{\partial\tau}{\partial\zeta},\\
H_{y} & = & \frac{\partial A_{x}}{\partial\zeta}-\frac{\partial A_{z}}{\partial\xi} & = & \frac{l'}{\sigma}\frac{\partial\tau}{\partial\zeta}-\frac{n'}{\sigma}\frac{\partial\tau}{\partial\xi},\\
H_{z} & = & \frac{\partial A_{y}}{\partial\xi}-\frac{\partial A_{x}}{\partial\eta} & = & \frac{m'}{\sigma}\frac{\partial\tau}{\partial\xi}-\frac{l'}{\sigma}\frac{\partial\tau}{\partial\eta},\\
E_{x} & = & -\frac{\partial\Phi}{\partial\xi}-\frac{1}{c}\frac{\partial A_{x}}{\partial\tau} & = & -\frac{cp'}{\sigma}\frac{\partial\tau}{\partial\xi}-\frac{l'}{c\sigma},\\
E_{y} & = & -\frac{\partial\Phi}{\partial\eta}-\frac{1}{c}\frac{\partial A_{y}}{\partial\tau} & = & -\frac{cp'}{\sigma}\frac{\partial\tau}{\partial\eta}-\frac{m'}{c\sigma},\\
E_{z} & = & -\frac{\partial\Phi}{\partial\zeta}-\frac{1}{c}\frac{\partial A_{z}}{\partial\tau} & = & -\frac{cp'}{\sigma}\frac{\partial\tau}{\partial\zeta}-\frac{n'}{c\sigma}.\end{array}\right\} \label{eq:Spawn}\end{equation}
In the above, Bateman's\begin{equation}
K\equiv\frac{\partial\sigma}{\partial\tau}=l'(x-\xi)+m'(y-\eta)+n'(z-\zeta)-c^{2}p'(t-\tau)=0\label{eq:K}\end{equation}
when the $(x,y,z,t)$ lie on the Eq.(\ref{eq:TangentPoint}) line
at $\tau$, because there $K=R(ll'+mm'+nn'-c^{2}pp')/cp=0$ by Eq.(\ref{eq:WaveCondition}).
The Eqs.(\ref{eq:Spawn}) simplify considerably when the Eq.(\ref{eq:gamma})
$\Gamma$ lies in a plane. Then, for zero Eq.(\ref{eq:sigma}) $\sigma$
the orbit's normal $\tau$ derivative vanishes, $\frac{\partial\tau}{\partial\zeta}=0$
when $\zeta(\tau)=0$ for example. If both $n$ and $n'$ vanish for
rays in the instanced orbit plane, the Eqs.(\ref{eq:Spawn}) spawned
fields have magnetic component normal to and electric fields in the
Eq.(\ref{eq:gamma}) $\Gamma$ plane. More general rays are possible.
In particular, rays with components normal to the orbit plane may
be found. But electric fields for these \cite[Figure 5]{Hannay} have
components normal to the orbit plane and, having longitudinal components,
are not even Maxwell light like. If they exist, their intensities
must be small, because only orbital plane polarization has been reported
\cite[Ref. 2]{Elder}.

Using Eqs.(\ref{eq:Spawn}) we can confirm that $\mathbf{E}\bullet\mathbf{H}=0$
and derive an expression for the Poynting vector $\mathbf{E}\times\mathbf{H}$.
Bateman says the Poynting vector is along the line described by Eq.(\ref{eq:TangentPoint}).
This must be since the singularity moves with the expanding Eq.(\ref{eq:R^2sphere})
sphere radius, but longitudinal field absence must be confirmed. Finally,
the Eq.(\ref{eq:Spawn}) fields must be shown to satisfy the Eq.(\ref{eq:MaxwellEqs})
Maxwell field relations. Bateman's work is incomplete, because he
never addresses these problems. Below, I show how these concerns can
restrict the possible fields and apply the result to synchrotron radiation.

To use the Eq.(\ref{eq:Spawn}) spawned fields in the Eq.(\ref{eq:MaxwellEqs})
Maxwell relations, I will consider the $(x,y,z,t)$ coordinate dependence
to be that contained in the denominator's Eq.(\ref{eq:sigma}) $\sigma$.
But after component differentiation $\sigma$ can be removed as serving
only as a space-time location indicator at the  singularity \emph{on
the expanding wave front}. This removal will be indicated by lower
case fields represented by transformations like  $h_{x}=\sigma H_{x}$.
With this notation the spawned fields must satisfy the following:\begin{equation}
\left.\begin{array}{ccc}
\boldsymbol{\nabla}\bullet\mathbf{H}=0 & \Rightarrow & lh_{x}+mh_{y}+nh_{z}=0\\
(\boldsymbol{\nabla}\times\mathbf{E}+\frac{1}{c}\frac{\partial\mathbf{H}}{\partial t})_{x}=0 & \Rightarrow & -cph_{x}+m\varepsilon_{z}-n\varepsilon_{y}=0\\
(\boldsymbol{\nabla}\times\mathbf{E}+\frac{1}{c}\frac{\partial\mathbf{H}}{\partial t})_{y}=0 & \Rightarrow & -cph_{y}+n\varepsilon_{x}-l\varepsilon_{z}=0\\
(\boldsymbol{\nabla}\times\mathbf{E}+\frac{1}{c}\frac{\partial\mathbf{H}}{\partial t})_{z}=0 & \Rightarrow & -cph_{z}+l\varepsilon_{y}-m\varepsilon_{x}=0\\
\boldsymbol{\nabla}\bullet\mathbf{E}=0 & \Rightarrow & l\varepsilon_{x}+m\varepsilon_{y}+n\varepsilon_{z}=0\\
(\boldsymbol{\nabla}\times\mathbf{H}-\frac{1}{c}\frac{\partial\mathbf{E}}{\partial t})_{x}=0 & \Rightarrow & cp\varepsilon_{x}+mh_{z}-nh_{y}=0\\
(\boldsymbol{\nabla}\times\mathbf{H}-\frac{1}{c}\frac{\partial\mathbf{E}}{\partial t})_{y}=0 & \Rightarrow & cp\varepsilon_{y}+nh_{x}-lh_{z}=0\\
(\boldsymbol{\nabla}\times\mathbf{H}-\frac{1}{c}\frac{\partial\mathbf{E}}{\partial t})_{z}=0 & \Rightarrow & cp\varepsilon_{z}+lh_{y}-mh_{x}=0\end{array}\right\} \label{eq:FieldConditions}\end{equation}

When written out using the Eqs.(\ref{eq:Spawn}) spawned fields, the
first four Eq.(\ref{eq:FieldConditions}) relations are homogeneous
expressions for the $l',m',n'$ and $cp'$ with coefficient determinant
\begin{equation}
\left|\begin{array}{cccc}
m\frac{\partial\tau}{\partial\zeta}-n\frac{\partial\tau}{\partial\eta} & n\frac{\partial\tau}{\partial\xi}-l\frac{\partial\tau}{\partial\zeta} & l\frac{\partial\tau}{\partial\eta}-m\frac{\partial\tau}{\partial\xi} & 0\\
0 & cp\frac{\partial\tau}{\partial\zeta}+\frac{n}{c} & -cp\frac{\partial\tau}{\partial\eta}-\frac{m}{c} & -m\frac{\partial\tau}{\partial\zeta}+n\frac{\partial\tau}{\partial\eta}\\
-cp\frac{\partial\tau}{\partial\zeta}-\frac{n}{c} & 0 & cp\frac{\partial\tau}{\partial\xi}+\frac{l}{c} & l\frac{\partial\tau}{\partial\zeta}-n\frac{\partial\tau}{\partial\xi}\\
cp\frac{\partial\tau}{\partial\eta}+\frac{m}{c} & -cp\frac{\partial\tau}{\partial\xi}-\frac{l}{c} & 0 & m\frac{\partial\tau}{\partial\xi}-l\frac{\partial\tau}{\partial\eta}\end{array}\right|.\label{eq:hCoefDeterminant}\end{equation}
Inspection shows that the $\boldsymbol{\nabla}\bullet\mathbf{H}$
expression is linearly dependent on the three $\boldsymbol{\nabla}\times\mathbf{E}$
expressions. When the Eq.(\ref{eq:gamma}) $\Gamma$ lies in a plane
containing the ray, only one $\boldsymbol{\nabla}\times\mathbf{E}$
expression remains, that giving the normal $\mathbf{H}$ field. When
written out using the Eqs.(\ref{eq:Spawn}) spawned fields, the last
four Eq.(\ref{eq:FieldConditions}) relations have the coefficient
determinant\begin{equation}
\left|\begin{array}{cccc}
-\frac{l}{c} & -\frac{m}{c} & -\frac{n}{c} & -l\frac{\partial\tau}{\partial\xi}-m\frac{\partial\tau}{\partial\eta}-n\frac{\partial\tau}{\partial\zeta}\\
-p-m\frac{\partial\tau}{\partial\eta}-n\frac{\partial\tau}{\partial\zeta} & m\frac{\partial\tau}{\partial\xi} & n\frac{\partial\tau}{\partial\xi} & -cp\frac{\partial\tau}{\partial\xi}\\
l\frac{\partial\tau}{\partial\eta} & -p-l\frac{\partial\tau}{\partial\xi}-n\frac{\partial\tau}{\partial\zeta} & n\frac{\partial\tau}{\partial\eta} & -cp\frac{\partial\tau}{\partial\eta}\\
l\frac{\partial\tau}{\partial\zeta} & m\frac{\partial\tau}{\partial\zeta} & -p-l\frac{\partial\tau}{\partial\xi}-m\frac{\partial\tau}{\partial\eta} & -cp\frac{\partial\tau}{\partial\zeta}\end{array}\right|.\label{eq:eCoefDeterminant}\end{equation}
In this case, inspection shows that the $\boldsymbol{\nabla}\bullet\mathbf{E}$
expression is linearly dependent on the three $\boldsymbol{\nabla}\times\mathbf{H}$
expressions and these allow \begin{equation}
\left.\begin{array}{cccccc}
 & m\frac{\partial\tau}{\partial\eta} & +n\frac{\partial\tau}{\partial\zeta} & =-p & \textrm{when} & \frac{\partial\tau}{\partial\xi}=0,\\
l\frac{\partial\tau}{\partial\xi} &  & +n\frac{\partial\tau}{\partial\zeta} & =-p & \textrm{when} & \frac{\partial\tau}{\partial\eta}=0,\\
l\frac{\partial\tau}{\partial\xi} & +m\frac{\partial\tau}{\partial\eta} &  & =-p & \textrm{when} & \frac{\partial\tau}{\partial\zeta}=0.\end{array}\right\} \label{eq:SpawnCondition}\end{equation}

As an example, consider the case where the primary singularity follows
the curve $\Gamma$ given by\begin{equation}
\begin{array}{ccccc}
x_{p}=\rho\cos\omega\tau, & y_{p}=\rho\sin\omega\tau, & z_{p}=0, & \textnormal{}{and} & t_{p}=\tau\end{array}\label{eq:ExampleGamma}\end{equation}
for constant $\rho$ and $\omega$. Using the functional designations
in Eq.(\ref{eq:gamma}), we find\begin{equation}
\begin{array}{ccc}
\tan\omega\tau=\frac{\eta}{\xi}, & \frac{\partial\tau}{\partial\xi}=-\frac{\eta}{v\rho}, & \frac{\partial\tau}{\partial\eta}=\frac{\xi}{v\rho}\end{array}\label{eq:exTauRelations}\end{equation}
when $v=\omega\rho$, $\xi(\tau)=\rho\cos(\omega\tau)$ and $\eta(\tau)=\rho\sin(\omega\tau)$.
 So, for $n=0$ we have the Eqs.(\ref{eq:SpawnCondition}) and (\ref{eq:WaveCondition})
to solve for $\frac{l}{cp}$ and $\frac{m}{cp}$. This gives\begin{equation}
\begin{array}{cc}
\frac{l_{\pm}}{cp}=\beta\sin(\omega\tau)\pm\sqrt{1-\beta^{2}}\cos(\omega\tau), & \frac{m_{\pm}}{cp}=-\beta\cos(\omega\tau)\pm\sqrt{1-\beta^{2}}\sin(\omega\tau)\end{array}\label{eq:exSpawnDirections}\end{equation}
where $\beta=\frac{v}{c}$. Using these, we can now calculate the
angle at the Eq.(\ref{eq:R^2sphere}) sphere center between the singularity
velocity on the ray $\mathbf{c}=(\frac{l}{p},\frac{m}{p},0)$ and
the singularity velocity on the Eq.(\ref{eq:gamma}) curve:\begin{equation}
\cos\delta=\frac{\mathbf{c}\bullet\mathbf{v}}{\sqrt{(\mathbf{c}\bullet\mathbf{c})(\mathbf{v}\bullet\mathbf{v})}}=\frac{\frac{l}{cp}\xi'+\frac{m}{cp}\eta'}{v}=-\beta.\label{eq:exCosSpawnAngle}\end{equation}
This recovers Eq.(\ref{eq:BatemanCondition}) and agrees with the
value given in \cite{Schwinger}, but I have found no published light
beam divergence measurements. Since Eq.(\ref{eq:PotentialDefs}) $p$
includes particle charge, changing its sign allows ray direction reversal.
 From Eqs.(\ref{eq:FieldConditions}), we find the fields:\begin{equation}
\begin{array}{ccc}
\varepsilon_{x}=-\frac{m_{\pm}}{cp}h_{z}, & \varepsilon_{y}=\frac{l\pm}{cp}h_{z}, & h_{z}=\frac{l\pm}{cp}\varepsilon_{y}-\frac{m_{\pm}}{cp}\varepsilon_{x}\end{array}.\label{eq:exHorizontalRayFields}\end{equation}
Clearly, the spawned $\boldsymbol{\varepsilon}$ field is normal to
the spawn direction, $\boldsymbol{\varepsilon}\bullet\boldsymbol{\mathbf{c}}=0.$
So the Poynting vector is in the spawn direction. When the directional
coefficients in the electric field components are held constant at
their spawn time values but $h_{z}$ is taken to be a periodic $t_{s}$
function, the secondary singularities will carry periodic fields along
the line given by\begin{equation}
\begin{array}{ccccc}
x_{s}=R\frac{l}{cp}+x_{p}, & y_{s}=R\frac{m}{cp}+y_{p}, & z_{s}=0 & \textnormal{}{and} & t_{s}=\frac{R}{c}+t_{p}\end{array}\label{eq:ExampleTangentPoint}\end{equation}
with polarization in the orbit plane in conformity with Eq.(\ref{eq:Spawn})
and as observed \cite[Ref. 2]{Elder}. To my knowledge, this prediction
is new.  The fields satisfy the Eq.(\ref{eq:MaxwellEqs}) Maxwell
equations when the periodic function argument is taken to be the retarded
time. This example shows that the Bateman scheme for  secondary singularity
association with light can give new, corroborative and verified results. 

To reinforce this conclusion, consider rays emitted at a fixed point
on the orbit described by Eqs.(\ref{eq:ExampleGamma}) and (\ref{eq:exTauRelations})
ignoring radial motion. In a plane normal to the orbit and tangent
to it at $x_{p}=\rho$ and $y_{p}=0$, the rays will be described
by Eq.(\ref{eq:ExampleTangentPoint}) with $l=0$ and $z_{s}=R\frac{n}{cp}$.
While $\frac{\partial\tau}{\partial\zeta}$ still vanishes, we now
have from Eq.(\ref{eq:exTauRelations}) the conditions $\frac{\partial\tau}{\partial\xi}=0$
and $\frac{\partial\tau}{\partial\eta}=\frac{1}{v}$. Using these
in Eq.(\ref{eq:Spawn}) shows that $h_{x}$, $\varepsilon_{y}$ and
$\varepsilon_{z}$ are the only nonzero field components with Eqs.(\ref{eq:FieldConditions})
requiring the relations\begin{equation}
\begin{array}{ccc}
\varepsilon_{y}=-\frac{n}{cp}h_{x}, & \varepsilon_{z}=\frac{m}{cp}h_{x}, & h_{x}=\frac{m}{cp}\varepsilon_{z}-\frac{n}{cp}\varepsilon_{y}\end{array}.\label{eq:VerticleRayFields}\end{equation}
Eqs.(\ref{eq:hCoefDeterminant}) and (\ref{eq:eCoefDeterminant})
show that satisfying the Eq.(\ref{eq:MaxwellEqs}) Maxwell relations
at the spawn point requires $\frac{p'}{p}=\frac{m'}{m}=\frac{n'}{n}$,
$p=-\frac{m}{v}$ and $n^{2}=(\frac{c^{2}}{v^{2}}-1)m^{2}$. Clearly
the Eq.(\ref{eq:VerticleRayFields}) $h_{x}$ is orthogonal to the
spawned electric field and both are orthogonal to the ray direction,
$(0,\frac{m}{cp},\frac{n}{cp})$; but as predicted above the polarization
has a component orthogonal to the orbit plane. As a final point, the
ray directions in planes between the orbit plane and the above orthogonal
plane can be examined. For planes touching the orbit at $x_{p}=\rho$
and $y_{p}=0$ with normal $\left(\cos\Theta,0,\sin\Theta\right)$,
included rays with direction $\left(\frac{l}{cp},\frac{m}{cp},\frac{n}{cp}\right)$
must satisfy the relation $l\cos\Theta+n\sin\Theta=0$. For the Eq.(\ref{eq:ExampleGamma})
orbit, Eq.(\ref{eq:SpawnCondition}) and Eq.(\ref{eq:WaveCondition})
give $\frac{m}{cp}=-\beta$ and $(\frac{l}{cp})^{2}=(1-\beta^{2})\sin^{2}\Theta$.
In these planes the angle between the ray and orbit direction is $\cos\delta=\left(\frac{l}{cp},\frac{m}{cp},\frac{n}{cp}\right)\bullet(0,1,0)$,
but again the polarization has a component orthogonal to the orbit
plane.  From this and \cite{Elder}, I conclude that, although electromagnetic
fields exist in all ray directions \cite{Hannay} for the Eq.(\ref{eq:PotentialDefs})
potentials, only Bateman rays from orbits for which $\mathbf{v}$
changes direction have significant intensity as required by Eq.(\ref{eq:Spawn}).
The implication is stark: Bateman rays enhance conventional electromagnetic
field descriptions.

Bateman has published other scheme applications. In his book \cite[Ch. VIII ]{Wave-motion},
Bateman applies his scheme to developing the electromagnetic field
for a simple moving singularity and by superposition to a moving doublet.
He also revisits the secondary singularity problem. The general procedure
is the same for all these examples. First, define a generalized potential
satisfying the homogeneous wave equation with singularities appropriate
for the problem. Then develop expressions for the electromagnetic
fields that are appropriate for the chosen generalized potential.
This method associates singularities with the electromagnetic field
to give the photon localization \cite{Dimitrova+} that has long eluded
explanation \cite{Keller}. For this reason, the method warrants further
evaluation with application specifically directed to describing familiar
light behaviors such as reflection, refraction and diffraction. The
last needs work \cite[pp.90-94]{Wave-motion}. But the logarithmic
generalized potential in \cite{Pairs} does describe ray bifurcation.
Requiring the potential function to satisfy a homogeneous wave equation
imposes constraints on the incoming and outgoing ray parameters. Matching
conditions at the bifurcation can be chosen to describe pair production,
Compton scattering or reflection with refraction. Bateman mentions
only the first possibility and none have been worked out. But with
the development presented here as a guide the work should be facilitated.
With this, sequential composition with \cite{Dempster} matching would
provide a purely mathematical explanation for  observed Mach-Zehnder
duality \cite{Jacques,Galvez}. The corpuscular response should be
easily obtained. Interference, however, is likely to require the wave
front singularities to be assigned nonzero spatial dimensions and
the path length difference to be less than the corpuscle coherence
length \cite[Sec. IV.C]{Gisin}. I leave these developments as an
exercise for the interested reader. Prior explanation attempts \cite[Sec. 5.]{Marshall+}
and \cite{Kaloyerou,Schneider}, while lacking Bateman localization,
 provide relevant considerations.

Clearly, our undergraduates should be introduced to Bateman's solution
to the wave/corpuscle duality problem on first exposure to the wave
equation. This exposure could be reinforced with subsequent laboratory
work \cite{Schneider,Sahyun}.\newpage{}


\begin{thebibliography}{10}
\bibitem{Thanks}I would like to thank an anonymous reviewer for comments
that confirm my introductory statement about Bateman's works on electromagnetism.
\end{thebibliography}
\end{document}